\documentclass[preprint,preprintnumbers,amsmath,amssymb]{revtex4}
\usepackage{epsfig}
\usepackage{graphicx}
\usepackage{dcolumn}
\usepackage{bm}
\usepackage{threeparttable}

\def\beq{\begin{equation}}
\def\eeq{\end{equation}}
\def\eeqn{\end{equation}}
\newcommand\iden{\leavevmode\hbox{\small1\normalsize\kern-.33em1}}

\newcommand{\bea} {\begin{eqnarray}}
\newcommand{\eea} {\end{eqnarray}}

\let\jnfont=\rm
\def\NPB#1,{{\jnfont Nucl.\ Phys.\ B }{\bf #1},}
\def\PLB#1,{{\jnfont Phys.\ Lett.\ B }{\bf #1},}
\def\EPJC#1,{{\jnfont Eur.\ Phys.\ Jour.\ C }{\bf #1},}
\def\PRD#1,{{\jnfont Phys.\ Rev.\ D }{\bf #1},}
\def\PRL#1,{{\jnfont Phys.\ Rev.\ Lett.\ }{\bf #1},}
\def\MPLA#1,{{\jnfont Mod.\ Phys.\ Lett.\ A }{\bf #1},}
\def\JPG#1,{{\jnfont J.\ Phys.\ G }{\bf #1},}
\def\CTP#1,{{\jnfont Commun.\ Theor.\ Phys.\ }{\bf #1},}
\def\JHEP#1,{{\jnfont JHEP \ }{\bf #1},}
\def\NPPS#1,{{\jnfont Nucl.\ Phys.\ Proc.\ Suppl.\ }{\bf #1},}
\def\CPC#1,{{\jnfont Comput.\ Phys.\ Commun.\ }{\bf #1},}
\def\CPL#1,{{\jnfont Chin.\ Phys.\ Lett. }{\bf #1},}
\def\APPB#1,{{\jnfont Acta\ Phys.\ Polon.\ B }{\bf #1},}
\def\PR#1,{{\jnfont Phys.\ Rept.\  }{\bf #1},}
\def\CHC#1,{{\jnfont Chin.\ Phys.\ C }{\bf #1},}

\def\lsim{\raise0.3ex\hbox{$<$\kern-0.75em\raise-1.1ex\hbox{$\sim$}}}
\def\gsim{\raise0.3ex\hbox{$>$\kern-0.75em\raise-1.1ex\hbox{$\sim$}}}

\begin{document}

\title{\ \\[10mm] A $\mu$-$\tau$-philic Higgs doublet confronted with the muon g-2, $\tau$ decays and LHC data}

\author{Lei Wang$^{1}$, Yang Zhang$^{2}$}
 \affiliation{$^1$ Department of Physics, Yantai University, Yantai
264005, P. R. China\\
$^2$ ARC Centre of Excellence for Particle Physics at the Tera-scale, School of Physics and Astronomy, Monash University, Melbourne, Victoria 3800, Australia
}


\begin{abstract}
In the framework of the two-Higgs-doublet model, one Higgs doublet may have the same interactions with fermions as the SM, 
and another Higgs doublet only 
has the $\mu$-$\tau$ LFV interactions. Assuming that the Yukawa matrices are real and symmetrical,
we impose various relevant theoretical and experimental constraints, and find that the excesses of muon $g-2$ and 
lepton flavour universality in the $\tau$ decays can be simultaneously explained in the region of small mass splittings between
the heavy CP-even Higgs and the CP-odd Higgs ($m_A > m_H$).
The multi-lepton event searches at the LHC can 
sizably reduce the mass ranges of extra Higgses, and $m_H$ is required to be larger than
560 GeV. 
\end{abstract}

\maketitle

\section{Introduction}
The muon anomalous magnetic moment $g-2$ has been
a long-standing puzzle since the announcement by the E821 experiment
in 2001~\cite{mug2-exp}. There is an almost $3.7\sigma$
discrepancy between the experimental value
and the prediction of the SM \cite{mug2-3.7}
\bea
\Delta a_\mu=a_\mu^{exp}-a_\mu^{SM}=(274\pm73)\times10^{-11}.
\eea 

The lepton flavor universality (LFU) in the $\tau$ decays is an excellent way
to probe new physics.
The HFAG collaboration reported three ratios from pure leptonic processes, and two ratios
from semi-hadronic processes, $\tau \to \pi/K \nu$ and $\pi/K \to \mu \nu$ \cite{tauexp}
\begin{eqnarray} \label{hfag-data}
&&
\left( g_\tau \over g_\mu \right) =1.0011 \pm 0.0015,~~
\left( g_\tau \over g_e \right) = 1.0029 \pm 0.0015,~~ \nonumber\\
&&
\left( g_\mu \over g_e \right) = 1.0018 \pm 0.0014, 
\left( g_\tau \over g_\mu \right)_\pi = 0.9963 \pm 0.0027,\nonumber\\
&&
\left( g_\tau \over g_\mu \right)_K = 0.9858 \pm 0.0071,
\end{eqnarray}
 where the ratios of $\left( g_\tau \over g_e \right)$ and $\left( g_\tau \over g_\mu \right)_K$ 
have approximate $2\sigma$ discrepancy from the SM.

As a simple extension of the SM, the lepton-specific two-Higgs-doublet model (2HDM) can
accommodate the muon $g-2$ anomaly by the contributions of two-loop Barr-Zee diagrams for a light CP-odd Higgs $A$ and large $\tan\beta$
 \cite{mu2h1-1,mu2h2,mu2h4,mu2h5,mu2h8,mu2h9,mu2h10,mu2h11,mu2h16,mu2h24,mu2h25,mu2h26}.
However, the tree-level diagram mediated by the charged Higgs gives negative contribution
to the decay $\tau\to \mu\nu\bar{\nu}$, which will raise the deviation of the LFU in
$\tau$ decays \cite{mu2h10,mu2h16,mueg2}. In addition, a scalar with the $\mu$-$\tau$ LFV interactions can accommodate the muon $g-2$ anomaly by
the contribution of one-loop diagrams \cite{0207302,10010434,150207824,151108544,151108880,160604408,190410908}.
Recently, Ref. \cite{190410908} showed that the excesses of muon $g-2$ and LFU in $\tau$ decays
can be simultaneously explained in the 2HDM in which one Higgs doublet has
 the same interactions with fermions as the SM, and another Higgs doublet only 
has the $\mu$-$\tau$ LFV interactions. In this paper, we focus on applying the ATLAS and 
CMS direct searches at the LHC to constrain the parameter space explaining the excesses of muon $g-2$ and LUF in $\tau$ decays. 

Our work is organized as follows. In Sec. II we recapitulate the
model. In Sec. III we discuss the muon $g-2$, LUF in $\tau$ decays, and other relevant constraints, and then
use the direct search limits at the LHC to constrain the model. 
Finally, we give our conclusion in Sec. IV.

\section{The 2HDM with $\mu$-$\tau$-philic Higgs doublet}
In Ref. \cite{190410908}, an inert Higgs doublet $\Phi_2$ is introduced to the SM
under an abelian discrete $Z_4$ symmetry, and the $Z_4$ charge assignment 
is shown in Table I.
\begin{table}
\caption{The $Z_4$ charge assignment.}
\label{tab:matter}
\begin{tabular}{|c||c|c|c||c|c|c||c|c|c||c|c|}
\hline
&~~$Q_L^{i}$~~&~~$U_R^i~~$&~~$D_R^i$~~&~~$L_L^e$~~&~~$L_L^\mu$~~&~~$L_L^\tau$~~&~~$e_R$~~&~~$\mu_R$~~&~~$\tau_R$~~&~~$\Phi_1$~~&~~$\Phi_2$~~\\ \hline
    ~~Z$_4$~~& 1         & 1       & 1       & 1          & $i$             & $-i$          & 1     & $i$        & $-i$     & $1$  & -1     \\ \hline
\end{tabular}
\end{table}
The scalar potential of $\Phi_2$ and $\Phi_1$ is
given as
\begin{eqnarray} \label{V2HDM} \mathrm{V} &=& Y_1
(\Phi_1^{\dagger} \Phi_1) + Y_2 (\Phi_2^{\dagger}
\Phi_2)+ \frac{\lambda_1}{2}  (\Phi_1^{\dagger} \Phi_1)^2 +
\frac{\lambda_2}{2} (\Phi_2^{\dagger} \Phi_2)^2  \nonumber \\
&&+ \lambda_3
(\Phi_1^{\dagger} \Phi_1)(\Phi_2^{\dagger} \Phi_2) + \lambda_4
(\Phi_1^{\dagger}
\Phi_2)(\Phi_2^{\dagger} \Phi_1)\nonumber \\
&&+ \left[\frac{\lambda_5}{2} (\Phi_1^{\dagger} \Phi_2)^2 + \rm
h.c.\right].
\end{eqnarray}
We focus on the CP-conserving case, and all
$\lambda_i$ are real. The two complex
scalar doublets can be written as
\begin{equation} \label{field}
\Phi_1=\left(\begin{array}{c} G^+ \\
\frac{1}{\sqrt{2}}\,(v+h+iG^0)
\end{array}\right)\,, \ \ \
\Phi_2=\left(\begin{array}{c} H^+ \\
\frac{1}{\sqrt{2}}\,(H+iA)
\end{array}\right). \nonumber
\end{equation}
The $\Phi_1$ field has the vacuum expectation value (VEV) $v$=246
GeV, and the VEV of $\Phi_2$ field is zero. We determine $Y_1$ by requiring the scalar
potential minimization condition.
\beq
Y_1=-\frac{1}{2}\lambda_1 v^2.
\eeq

The $G^0$ and $G^+$ are the Nambu-Goldstone bosons which are eaten by the gauge bosons. 
The $H^+$ and $A$ are the mass eigenstates of the charged Higgs boson and
CP-odd Higgs boson.  
 Their masses are given as
\beq \label{masshp}
 m_{H^\pm}^2  = Y_2+\frac{\lambda_3}{2} v^2, ~~m_{A}^2  = m_{H^\pm}^2+\frac{1}{2}(\lambda_4-\lambda_5) v^2.
 \eeq
The two CP-even Higgses $h$ and $H$ are mass eigenstates, and there is no mixing between them.
In this paper, the light CP-even Higgs $h$ is taken as the SM-like Higgs. Their masses are given as
\beq \label{massh}
 m_{h}^2  = \lambda_1 v^2\equiv (125~{\rm GeV })^2, ~~m_{H}^2  = m_{A}^2+\lambda_5 v^2.
 \eeq

The masses of fermions are obtained from the Yukawa interactions with $\Phi_1$,
 \beq \label{yukawacoupling} - {\cal L} = y_u\overline{Q}_L \,
\tilde{{ \Phi}}_1 \,U_R + y_d\overline{Q}_L\,{\Phi}_1 \, D_R +  y_\ell\overline{L}_L \, {\Phi}_1
\, E_R + \mbox{h.c.}, \eeq
where $\widetilde\Phi_{1}=i\tau_2 \Phi_{1}^*$, $Q_L^T=(u_{Li}\,,d_{Li})$, $L_L^T=(\nu_{Li}\,,\ell_{Li})$ 
with $i$ being generation indices. $U_R$, $D_R$, and $E_R$ denote the three generation right-handed fields of
the up-type quark, down-type quark, and charged lepton.
According to the $Z_4$ charge assignment in Table I, the $Z_4$ symmetry allows the quark Yukawa matrix $y_u$ ($y_d$)
to be not diagonalized, and requires the lepton Yukawa matrix $y_\ell$ to be diagonal.
Therefore, the quark fields ($Q_L$, $U_R$, $D_R$) are the interaction eigenstates, 
and the lepton fields ($L_L$, $E_R$) are mass eigenstates.

The $Z_4$ symmetry allows $\Phi_2$ to have $\mu$-$\tau$ interactions \cite{190410908},
\bea\label{lepyukawa2}
- {\cal L}_{LFV} &=&  \sqrt{2}~\rho_{\mu\tau} \,\overline{L^\mu_{L}} \, {\Phi}_2
\,\tau_R  \, + \sqrt{2}~\rho_{\tau\mu}\, \overline{L^\tau_{L}} \, {\Phi}_2
\,\mu_R \, + \, \mbox{h.c.}\,. \eea
From Eq. (\ref{lepyukawa2}), we can obtain $\mu$-$\tau$ LFV couplings of extra Higgses ($H$, $A$, and $H^\pm$).
We assume that the Yukawa matrix of $\Phi_2$ is CP-conserving, namely that $\rho_{\mu\tau}$ and $\rho_{\tau\mu}$ 
are real and $\rho_{\mu\tau}=\rho_{\tau\mu}\equiv\rho$.

At the tree-level, the light CP-even Higgs $h$ has the same couplings to fermions and gauge boson as the SM, and 
the $\mu$-$\tau$ LFV coupling of $h$ is absent. The Yukawa couplings of $H$, $A$, and $H^\pm$ are $\mu$-$\tau$-philic,
and they have no other Yukawa couplings. The neutral Higgses $A$ and $H$ have no cubic interactions with $ZZ, ~WW$.

\section{Muon $g-2$, LUF in $\tau$ decays, LHC data, and relevant constraints}\label{constraints}
\subsection{Numerical calculations}
In our calculations, we take $\lambda_2$, $\lambda_3$, $m_h$, $m_H$, $m_A$ and $m_{H^\pm}$ as the input parameters, which
can determine the values of $\lambda_1$, $\lambda_5$ and $\lambda_4$ from Eqs. (\ref{masshp}, \ref{massh}).
 $\lambda_2$ controls the quartic couplings of extra Higgses, and does not affect the observables considered in our paper.
Therefore, we simply take $\lambda_2=\lambda_1$. $\lambda_3$ is adjusted to satisfy the theoretical constraints. We fix
$m_h=125$ GeV, and scan over several key parameters in the following ranges:
\bea
&&300~{\rm GeV}<m_H<800~{\rm GeV},~m_H < m_A=m_{H^{\pm}} < m_H + 200~ {\rm GeV},\nonumber\\
&&0.1 < \rho <1.0.
\label{range}
\eea

At the tree-level, the SM-like Higgs has the same couplings to the SM particles as the SM, and no exotic decay mode for 
such Higgs mass spectrum. The masses of extra Higgses are beyond the exclusion range of the searches for 
the neutral and charged Higgs at the LEP.
Because the extra Higgses have no couplings to quarks, 
the bounds from meson observables can be safely neglected.

In our calculation, we consider the following observables and constraints:

\begin{itemize}
\item[(1)] Theoretical constraints and precision electroweak data. The $\textsf{2HDMC}$ \cite{2hc-1}
is employed to implement the theoretical
constraints from the vacuum stability, unitarity and
coupling-constant perturbativity, and calculated 
the oblique parameters ($S$, $T$, $U$). Adopting the recent fit results in Ref. \cite{pdg2018}, we use the following 
values of $S$, $T$, $U$,
\beq
S=0.02\pm 0.10, ~~T=0.07\pm 0.12,~~U=0.00 \pm 0.09. 
\eeq
The correlation coefficients are given by
\beq
\rho_{ST} = 0.92, ~~\rho_{SU} = -0.66, ~~\rho_{TU} = -0.86.
\eeq
The oblique parameters favor that one of $H$ and $A$ has a small mass splitting from $H^\pm$, and therefore 
we simply take $m_A=m_{H^{\pm}}$ in this paper.

\item[(2)] Muon $g-2$. The model contributes to the muon $g-2$ through the one-loop diagrams involving 
the $\mu$-$\tau$ LFV coupling of $H$ and $A$ \cite{10010434}, 
\bea
  \delta a_{\mu}=\frac{m_\mu m_\tau \rho^2}{8\pi^2}
  \left[\frac{(\log\frac{m_H^2}{m_\tau^2}-\frac{3}{2})}{m_H^2}
  -\frac{\log(\frac{m_A^2}{m_\tau^2}-\frac{3}{2})}{m_A^2}
\right].
  \label{mua1}
\eea
From Eq. (\ref{mua1}), the model can give a positive contribution to the muon $g-2$ for $m_A>m_H$. This is reason why
we scan over the parameter space of $m_A > m_H$.

\item[(3)] Lepton universality in the $\tau$ decays.
The HFAG collaboration reported three ratios from pure leptonic processes,  
\begin{eqnarray} 
&&
\left( g_\tau \over g_\mu \right)^2 \equiv \bar{\Gamma}(\tau\to e
\nu\bar{\nu})/\bar{\Gamma}(\mu\to e \nu\bar{\nu}), \nonumber\\
&&
\left( g_\tau \over g_e \right)^2  \equiv \bar{\Gamma}(\tau\to \mu
\nu\bar{\nu})/\bar{\Gamma}(\mu\to e \nu\bar{\nu}), \nonumber\\
&&
\left( g_\mu \over g_e \right)^2  \equiv \bar{\Gamma}(\tau\to \mu
\nu\bar{\nu})/\bar{\Gamma}(\tau\to e \nu\bar{\nu}),
\end{eqnarray}
and two ratios
from semi-hadronic processes, $\tau \to \pi/K \nu$ and $\pi/K \to \mu \nu$ \cite{tauexp}.
The values of five ratios are given in Eq. (\ref{hfag-data}).
Here $\bar{\Gamma}$ denotes the partial width
normalized to its SM value. The correlation matrix for the above five observables is
\begin{equation} \label{hfag-corr}
\left(
\begin{array}{ccccc}
1 & +0.53 & -0.49 & +0.24 & +0.12 \\
+0.53  & 1     &  + 0.48 & +0.26    & +0.10 \\
-0.49  & +0.48  & 1       &   +0.02 & -0.02 \\
+0.24  & +0.26  & +0.02  &     1    &     +0.05 \\
+0.12  & +0.10  & -0.02  &  +0.05  &   1 
\end{array} \right) .
\end{equation}
In this model,
\begin{eqnarray} \label{tau-loop}
&&\bar{\Gamma}(\tau\to \mu \nu\bar{\nu})= (1+\delta_{\rm loop}^\tau)^2~(1+\delta_{\rm loop}^\mu)^2+\delta_{\rm tree},\nonumber\\
&&\bar{\Gamma}(\tau\to e \nu\bar{\nu})= (1+\delta_{\rm loop}^\tau)^2,\nonumber\\
&&\bar{\Gamma}(\mu\to e \nu\bar{\nu})= (1+\delta_{\rm loop}^\mu)^2.
\end{eqnarray}
Where $\delta_{\rm tree}$ can give a positive correction to $\tau\to \mu \nu\bar{\nu}$, and is from the tree-level diagram
mediated by the charged Higgs,
\beq
\delta_{\rm tree}=4\frac{m_W^4\rho^4}{g^4 m_{H^{\pm}}^4}.
\label{delta-tree}
\eeq
$\delta_{\rm loop}^\tau$ and $\delta_{\rm loop}^\mu$ denote the corrections to vertices $W\bar{\nu_{\tau}}\tau$ and
$W\bar{\nu_{\mu}}\mu$, respectively, which are from the one-loop diagrams involving $H$, $A$, and $H^\pm$. Since we
take $X_{\mu\tau}=X_{\tau\mu}$ for the lepton Yukawa matrix, and therefore $\delta_{\rm loop}^\tau=\delta_{\rm loop}^\mu$. Following 
results of \cite{mu2h10,mu2h16,190410908},
\beq
\delta_{\rm loop}^\tau=\delta_{\rm loop}^\mu={1 \over 16 \pi^2} {\rho^2} 
\left[1 + {1\over4} \left( H(x_A) +  H(x_H) \right)
\right]\,, 
\eeq
where $H(x_\phi) \equiv \ln(x_\phi) (1+x_\phi)/(1-x_\phi)$ with $x_\phi=m_\phi^2/m_{H^{\pm}}^2$.

In the model, 
\beq\left( g_\tau \over g_\mu \right)_\pi=\left( g_\tau \over g_\mu \right)_K =\left( g_\tau \over g_\mu \right).
\eeq
We perform $\chi^2_\tau$ calculation for the five observables. The covariance matrix constructed from the data of Eq. (\ref{hfag-data})
and Eq. (\ref{hfag-corr}) has a vanishing eigenvalue, and the corresponding degree is removed in our calculation.

\item[(4)] Lepton universality in the $Z$ decays. The measured values of the ratios of the leptonic $Z$ decay
branching fractions are given as \cite{zexp}:
\begin{eqnarray} \label{lu-zdecay}
{\Gamma_{Z\to \tau^+ \tau^- }\over \Gamma_{Z\to e^+ e^- }} &=& 1.0019 \pm 0.0032,\\
{\Gamma_{Z\to \mu^+ \mu^-}\over \Gamma_{Z\to e^+ e^- }} &=& 1.0009 \pm 0.0028,
\end{eqnarray}
with a correlation of $+0.63$. 
The model can give corrections to the widths of $Z\to \tau^+\tau^-$ and $Z\to \mu^+\mu^-$
through the one-loop diagrams involving the extra Higgs bosons. The quantities of Eq. (\ref{lu-zdecay}) are calculated in the model
 as \cite{mu2h10,mu2h16,190410908}
\beq 
{\Gamma_{Z\to \tau^+ \tau^- }\over \Gamma_{Z\to e^+ e^- }} 
\approx 1.0+ {2 g_L^e{\rm Re}(\delta g^{\rm loop}_L)+ 2 g_R^e{\rm Re}(\delta g^{\rm loop}_R) \over {g_L^e}^2 + {g_R^e}^2 }\,.
\,
\eeq
where the SM value $g_L^e=-0.27$ and $g_R^e=0.23$. $\delta g^{\rm loop}_L$ and $\delta g^{\rm loop}_R$
are from the one-loop corrections, which are given as 
\begin{eqnarray} \label{dgLR}
\delta g^{\rm loop}_L &=& {1\over 16\pi^2} \rho^2 \,
\bigg\{
 -{1\over2} B_Z(r_A)- {1\over2} B_Z(r_H) -2 C_Z(r_A, r_H)
   \nonumber \\
&&  + s_W^2 \left[ B_Z(r_A) + B_Z(r_H) + \tilde C_Z(r_A) + \tilde C_Z(r_H) \right] \bigg\} 
\,,\\ 
\delta g^{\rm loop}_R &=& {1\over 16\pi^2} \rho^2 \,
\bigg\{ 2 C_Z(r_A, r_H) - 2 C_Z(r_{H^\pm}, r_{H^\pm}) 
+ \tilde C_Z(r_{H^\pm}) 
\nonumber \\
&& - {1\over2} \tilde C_Z(r_A)  - {1\over2} \tilde C_Z(r_H)      
 +  s_W^2 \left[ B_Z(r_A) + B_Z(r_H) + 2 B_Z(r_{H^\pm})\right.
\nonumber \\
&& \left. +\tilde C_Z(r_A) + \tilde C_Z(r_H) + 4C_Z(r_{H^\pm},r_{H^\pm}) \right] \bigg\} 
\,,
\end{eqnarray}
where $r_\phi = m_\phi^2/m_Z^2$ with $\phi=A,H, H^\pm$, and 
\begin{eqnarray} 
\label{loopftn}
B_Z(r) &=& -{\Delta_\epsilon \over 2} -{1\over4} + {1\over2} \log(r) \,,\\ 
C_Z(r_1,r_2) &=& {\Delta_\epsilon \over4} -{1\over2} \int^1_0 d x \int^x_0 d y\,
\log[ r_2 (1-x) + (r_1 -1) y + x y] \,,\\ 
\tilde C_Z(r) &=& {\Delta_\epsilon \over2}+{1\over2}  - r\big[1+\log(r) \big]
+r^2 \big[ \log(r) \log(1+r^{-1}) \nonumber\\
&&
-{\rm Li_2}(-r^{-1}) \big] -{i \pi\over2}
\left[ 
	1 - 2r + 2r^{2}\log(1+r^{-1})
\right].
\end{eqnarray}

Due to $X_{\mu\tau}=X_{\tau\mu}$, we can obtain
\beq
{\Gamma_{Z\to \mu^+ \mu^-}\over \Gamma_{Z\to e^+ e^- }}={\Gamma_{Z\to \tau^+ \tau^- }\over \Gamma_{Z\to e^+ e^- }}.
\eeq

(5) The exclusions from the ATLAS and CMS searches at the LHC.
The extra Higgs bosons are dominantly produced at the LHC via
the following electroweak processes:
\begin{align}
pp\to & W^{\pm *} \to H^\pm A, \label{process1}\\
pp\to &       Z^* \to HA, \label{process2}\\
pp\to & W^{\pm *} \to H^\pm H, \label{process3}\\
pp\to & Z^*/\gamma^* \to H^+H^-. \label{process4}
\end{align}

For small mass splitting among $H$, $A$, and $H^{\pm}$, the dominant decay modes of these Higgses are
\beq
H\to \tau^{\pm}\mu^{\mp},~~~A\to \tau^{\pm}\mu^{\mp},~~~H^\pm\to \tau^\pm \nu_\mu, \mu^\pm \nu_\tau.
\label{decay-taumu}
\eeq
When $m_A$ and $m_{H^{\pm}}$ are much larger than $m_H$, the following exotic decay modes will open with $m_A=m_{H^\pm}$, 
\beq
A\to HZ, ~~~~H^\pm \to H W^\pm. 
\eeq

In order to restrict the productions of the above processes at the LHC for our model, 
we perform simulations for the samples 
using \texttt{MG5\_aMC-2.4.3}~\cite{Alwall:2014hca} 
with \texttt{PYTHIA6}~\cite{Torrielli:2010aw} and 
\texttt{Delphes-3.2.0}~\cite{deFavereau:2013fsa}, and adopt the constraints 
from all the analysis for the 13 TeV LHC in 
version \texttt{CheckMATE 2.0.26}~\cite{Dercks:2016npn}. Besides, the latest multi-lepton searches for 
electroweakino~\cite{Sirunyan:2017zss,Sirunyan:2017qaj,Sirunyan:2017lae,Sirunyan:2018ubx,Aaboud:2017nhr} implemented in Ref.~\cite{Pozzo:2018anw} and the ATLAS search for direct stau production with 139 fb$^{-1}$ 13 TeV events~\cite{ATLAS:2019ucg} are also taken into consideration.

\end{itemize}

\begin{figure}[tb]
  \epsfig{file=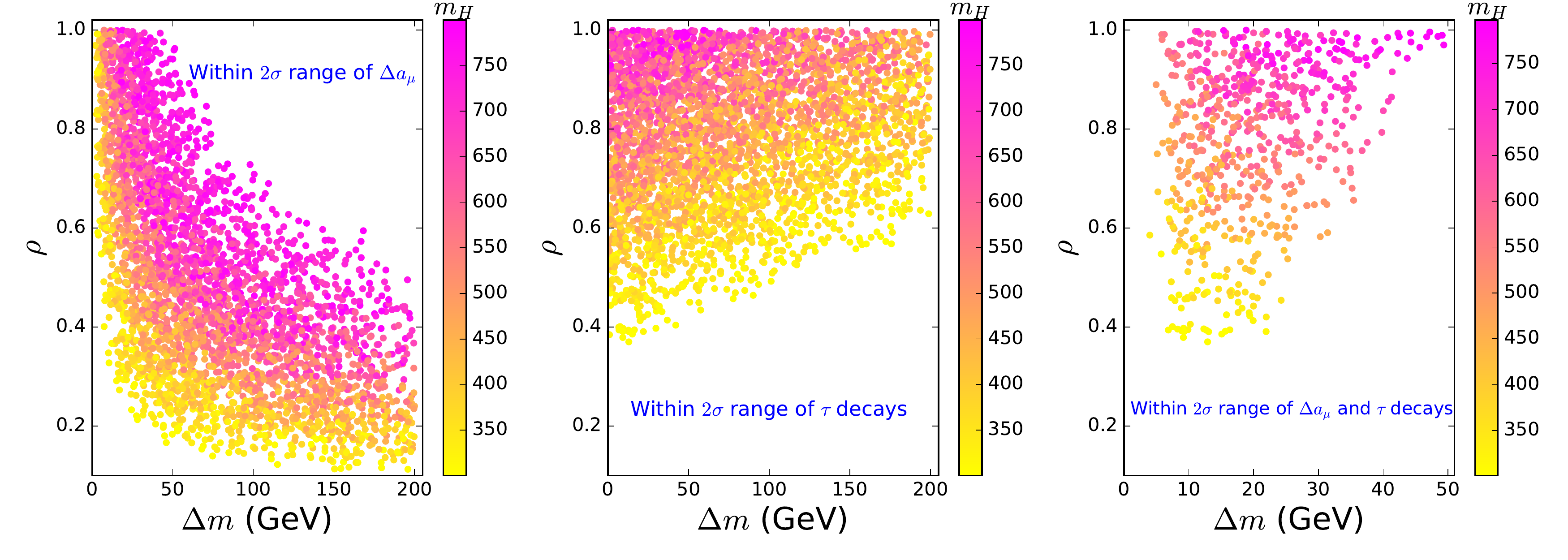,height=6.2cm}
  \epsfig{file=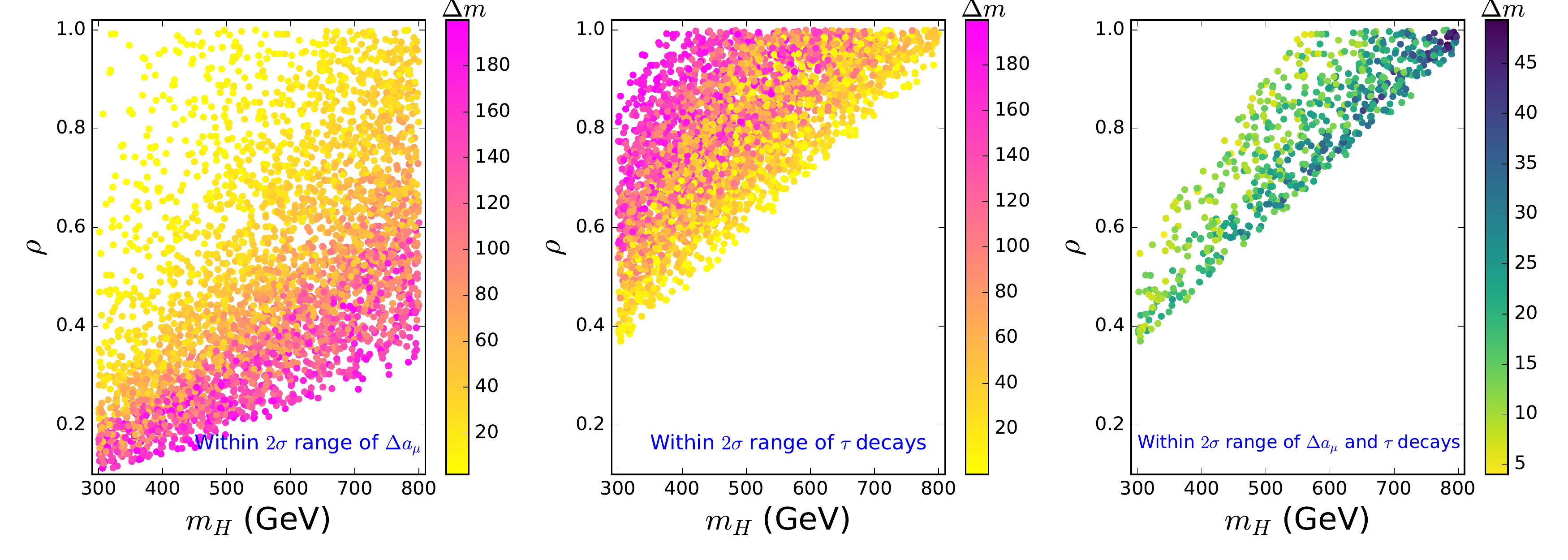,height=6.2cm}
\vspace{-0.5cm} \caption{The samples within $2\sigma$ ranges of $\Delta a_\mu$ (left panel), LUF in the $\tau$ decays (middle panel), and both
$\Delta a_\mu$ and LUF in the $\tau$ decays (right panel). All the samples satisfy the constraints of the theory, oblique parameters and $Z$ decays.
Here $\Delta m\equiv m_A (m_{H^\pm})-m_H$.}
\label{mg2tau}
\end{figure}

\subsection{Results and discussions}
We find that the constraints from theory, oblique parameters and $Z$ decays can be easily satisfied 
in the parameter space taken in this paper. The allowed ranges of $m_H$, $m_A$, $m_{H^{\pm}}$, and $\rho$
are not reduced by the those constraints. Therefore, we won't show their results in the following discussions.

After imposing the constraints of the theory, the
oblique parameters, and $Z$ decays, in Fig. \ref{mg2tau} we show the surviving samples which are consistent with
$\Delta a_\mu$ and $\tau$ decays at $2\sigma$ level. The contributions of $H$ and $A$ to $\Delta a_\mu$ are respectively 
positive and negative, and a large mass splitting between $m_A$ and $m_H$
can produce sizable corrections to $\Delta a_\mu$. Thus, with an increase of $\Delta m$, 
a relative small $\rho$ can make $\Delta a_\mu$ to be within $2\sigma$ range of experimental data, as shown in upper-left 
panel in Fig. \ref{mg2tau}.

The experimental value of $\left( g_\tau \over g_e \right)$ has about $2\sigma$ positive deviation from the SM prediction.
Enhancement of $\Gamma(\tau\to \mu \nu\bar{\nu})$ can provide a better fit. According to Eq. (\ref{tau-loop}), 
the $\delta_{\rm tree}$ term can enhance $\Gamma(\tau\to \mu \nu\bar{\nu})$, and favor $\rho$ to increase with $m_{H^\pm}$, see
Eq. (\ref{delta-tree}). Therefore, the upper-middle panel shows that the experimental data of LFU in the $\tau$ decays
 favor $\rho$ to increase with $\Delta m$. Because of the opposite relationship between $\rho$ and $\Delta m$, 
the excesses of $\Delta a_\mu$ and $\tau$ decays can be simultaneously explained in a narrow region of $\rho$ and $\Delta m$.
As shown in the upper-right panel, $\Delta m < 50$ GeV and $\rho>$ 0.36 are required, and $\rho$ is favored to increase with $\Delta m$.

The lower panels of Fig. \ref{mg2tau} show that the experimental data of $\Delta a_\mu$ and $\tau$ decays favor $\rho$ to increase with $m_H$. 
The lower-right panel shows that the excesses of $\Delta a_\mu$ and $\tau$ decays can be simultaneously explained in the
range of 300 GeV $<m_H<$ 800 GeV, and the corresponding $\rho$ is imposed upper and lower bounds. 
Taking $m_H=500$ GeV for an example, the excesses of $\Delta a_\mu$ and LUF in the $\tau$ decays can be simultaneously explained for 
$0.6 <\rho< 0.9$. For $m_H=500$ GeV and $\rho<$ 0.6, $\Delta a_\mu$ can be explained, but the $\tau$ decays can
not be accommodated. For $m_H=500$ GeV and $\rho>$ 0.9, $\Delta a_\mu$ and the $\tau$ decays can be respectively 
explained. However, the former favors a small $\Delta m$ and the latter favors a large $\Delta m$, which leads that
the two anomalies can not be simultaneously explained for $m_H=500$ GeV and $\rho>$ 0.9.

\begin{figure}[tb]
  \epsfig{file=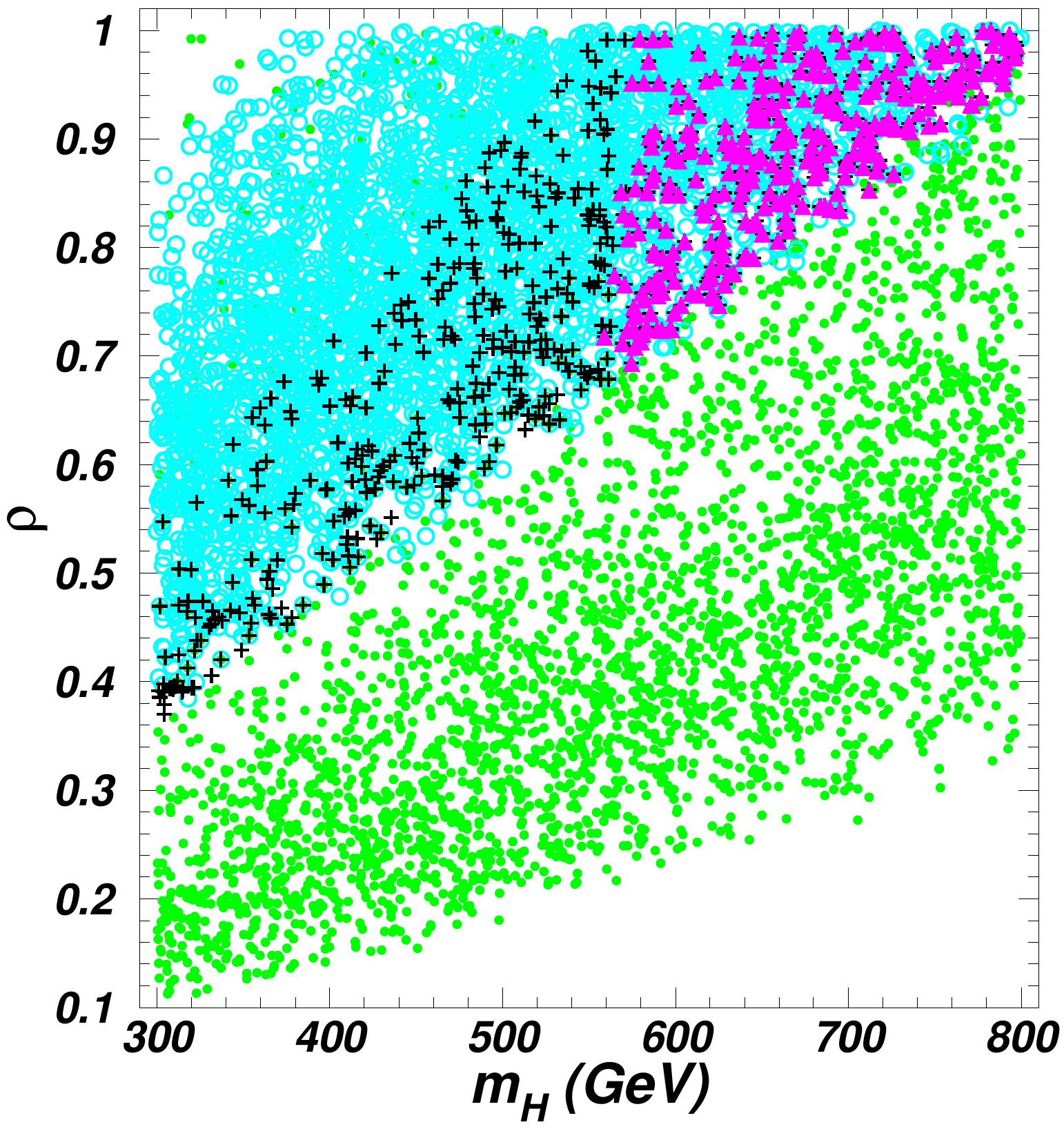,height=7.5cm}
  \epsfig{file=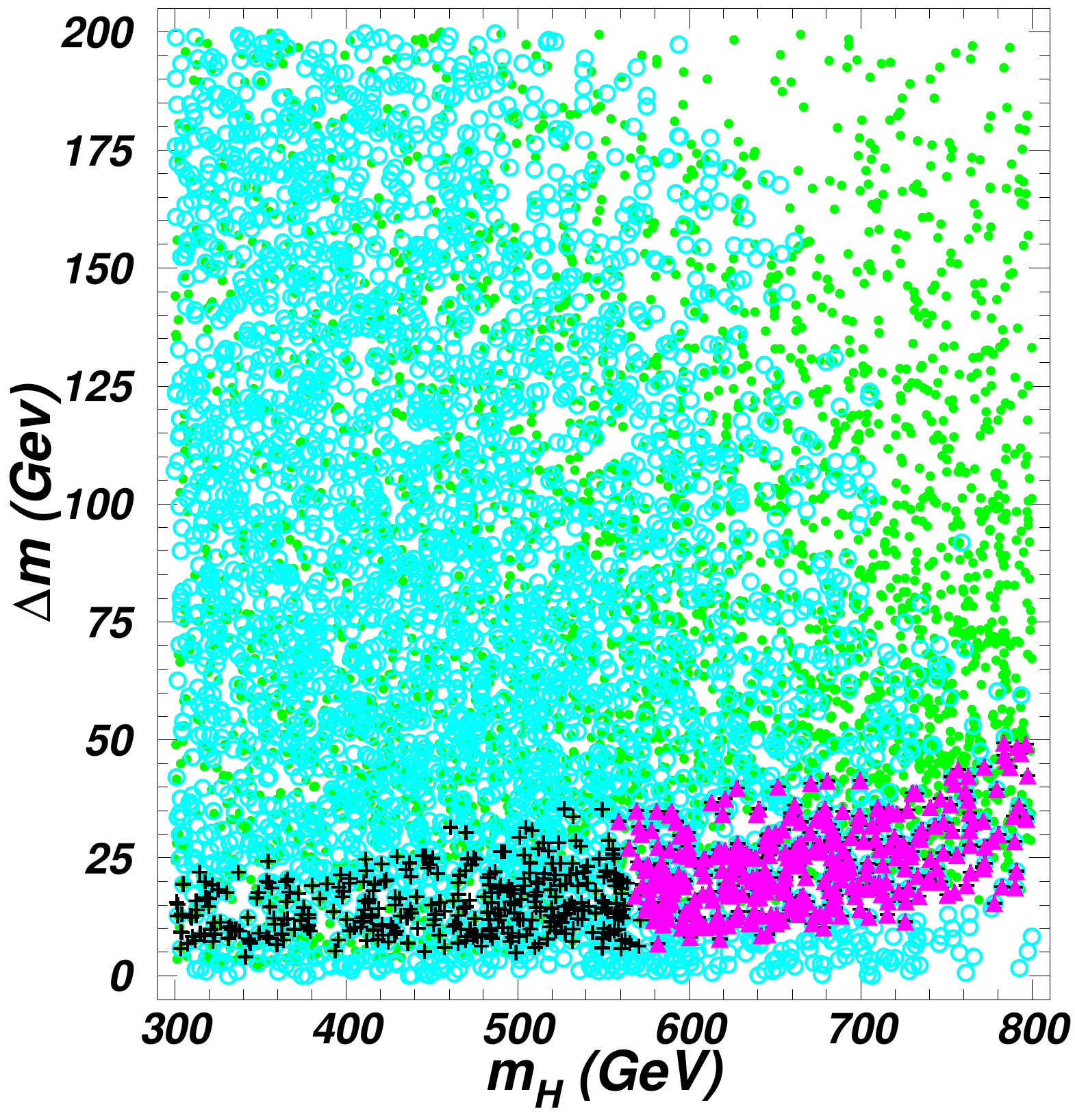,height=7.5cm}
\vspace{-0.5cm} \caption{The surviving samples on the planes
of $m_H$ versus $\rho$ and $m_H$ versus $\Delta m$. All the samples satisfy the constraints of the theory,
 oblique parameters and $Z$ decays. In addition, the bullets (green) within the $2\sigma$ ranges of muon $g-2$ and
 the circles (blue) within the $2\sigma$ ranges of LUF in the $\tau$ decays. The pluses (black) and triangles (purple) 
are within the $2\sigma$ ranges of both muon $g-2$ and LUF in the $\tau$ decays, and the former are excluded
 by the constraints of the direct searches at the LHC, while the latter are allowed.}
\label{lhc}
\end{figure}

After imposing the constraints of the direct searches at the LHC, those samples of Fig. \ref{mg2tau}
are projected on the planes of $m_H$ versus $\rho$ and $m_H$ versus $\Delta m$, as shown in Fig. \ref{lhc}.
Since the excesses of $\Delta a_\mu$ and $\tau$ decays require the mass splitting between $m_A$ ($m_{H^\pm}$) and $m_H$
 to be smaller than 50 GeV, $H$, $A$ and $H^\pm$ will dominantly decay into $\tau\mu$, $\tau\nu_\mu$, and $\mu\nu_\tau$. 
The direct searches at the LHC exclude region of $m_H< 560$ GeV, and the corresponding 
$\rho$ is required to be larger than 0.68. Since $\Delta m$ is such small, it hardly affects the excluded region.

\begin{figure}[tb]
  \epsfig{file=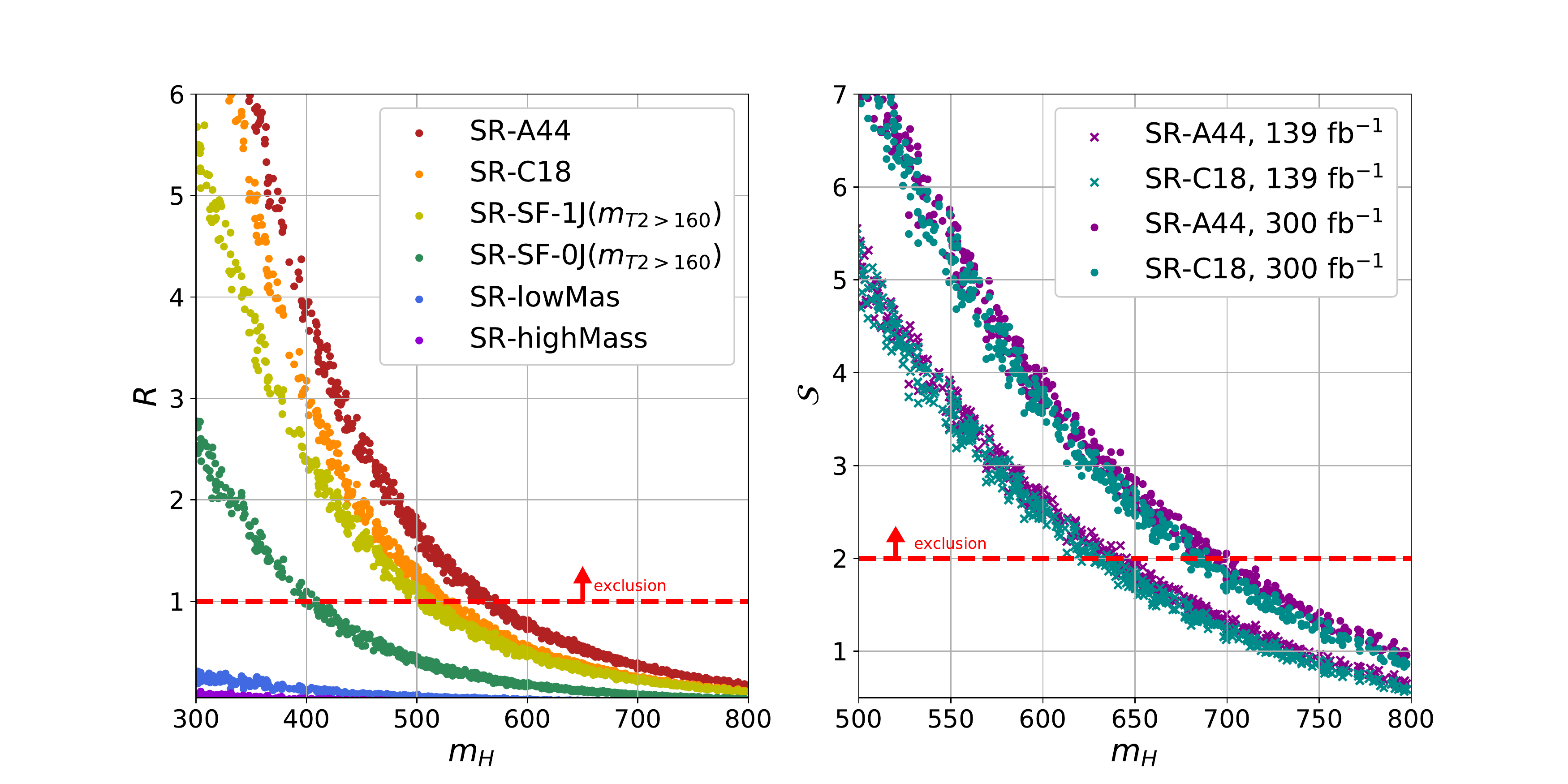,height=8.5cm}
\vspace{-0.5cm} \caption{Left panel: the ratio $R$ of event yields in the signal regions to the corresponding 95\% experimental limit for the surviving samples. The red and orange dots represent the signal regions \texttt{SR-A44} and \texttt{SR-C18} of the multilepton search with 35.9 fb$^{-1}$ LHC data~\cite{Sirunyan:2017lae}. The yellow and green dots stand the signal regions \texttt{SR-SF-1J} ($m_{T2}>160$~GeV) and \texttt{SR-SF-0J} ($m_{T2}>160$~GeV) of the sleptons search of two leptons with 139 fb$^{-1}$ LHC data~\cite{ATLAS:2019cfv}. The green and blue dots represent the signal regions \texttt{SR-lowMass} and \texttt{SR-highMass} of the stau search with 139 fb$^{-1}$ LHC data~\cite{ATLAS:2019ucg}. Right panel: the estimated expected significance $\mathcal{S}$ for the surviving sample with high integrated luminosity. The purple and cyan dots stand the signal regions \texttt{SR-A44} and \texttt{SR-C18} with 139 fb$^{-1}$ integrated luminosity data (crosses) and 300 fb$^{-1}$ integrated luminosity data (dots).}
\label{lhc-R}
\end{figure}

For the excluded samples, the most sensitive experimental analysis is the CMS search for electroweak production of charginos and neutralinos in multilepton final states~\cite{Sirunyan:2017lae} at 13 TeV LHC with 35.9 fb$^{-1}$ integrated luminosity data. In this analysis, hundreds of signal region bins are designed, of which the \texttt{SR-A44} and \texttt{SR-C18} provide strongest constraints on our samples. In the \texttt{SR-A44}, events are selected on the condition that contains three leptons forming at least one opposite-sign same-flavor (OSSF) pair, $M_{\ell\ell}>105$ GeV, $p^{\rm miss}_T>200$ GeV and $M_T>160$ GeV. Here $M_{\ell\ell}$ is the invariant mass of the OSSF dilepton pair, $p^{\rm miss}_T$ stands the missing transverse momentum, and $M_T=\sqrt{2 p^{\rm miss}_T p^{\ell}_T[1-\cos(\Delta \phi)]}$ is the transverse mass computed with respect to the third lepton in the event. The \texttt{SR-C} is built from events with two $e$ or $\mu$ forming an OSSF pair and a hadronic decay tau lepton $\tau_h$. The bin \texttt{SR-C18} requires $p^{\rm miss}_T>200$ GeV, $|M_{\ell\ell}-m_Z|>15$ GeV and $M_{T2}>100$ GeV. The two-lepton transverse mass $M_{T2}$ \cite{Lester:1999tx,Barr:2003rg} is computed with the OSSF pair of light leptons.
The main contributions of our samples to the bins are from processes in Eq.~(\ref{process1}) and Eq. (\ref{process3}) with one or two of the $\tau$s decaying hadronically. Thus, the exclusion power decreases gently with heavier $m_{H}$ and $M_A$ due to the smaller production rates for the processes, as shown in the left panel of Fig.~\ref{lhc-R}, where the $y$-axis $R$ stands the ratio of event yields in the signal region to the corresponding 95\% experimental limit.

We also adopt the ATLAS searches with 139 fb$^{-1}$ integrated luminosity data for slepton pair production~\cite{ATLAS:2019cfv} and direct stau pair production~\cite{ATLAS:2019ucg} by implementing them in \texttt{CheckMATE 2.0.26}. Although the integrated luminosity is much higher than that of the multilepton search~\cite{Sirunyan:2017lae}, the exclusion powers of the slepton search, which requires exactly two light leptons in final state, and the stau search are weaker than the above constraints. We can see from the left panel of Fig.~\ref{lhc-R} that the most sensitive signal region in the slepton search is the \texttt{SR-SF-1J} ($m_{T2}>160$~GeV), which can exclude samples of $m_H< 501$ GeV. In the \texttt{SR-SF-1J} ($m_{T2}>160$~GeV), events are required to have exactly two light flavour leptons and exactly one jet. Thus the dominated processes for this signal region are Eq.~(\ref{process1}) and Eq.~(\ref{process3}), where the $H^{\pm}$ provides one $\mu$ and $A$ or $H$ contributes one $\mu$ and one $\tau$ jet. Note that this bound, $m_H> 501$ GeV, is weaker than the estimated limit, $m_{\phi}>700$ GeV, adopted in Ref.~\cite{190410908}, which is the slepton mass bound in the massless neutralino limit \cite{ATLAS:2019cfv}. The detailed explanations are given in
the Appendix.

As for the stau search~\cite{ATLAS:2019ucg}, it can not give constraints to any sample with $m_H>300$ GeV. It is because that both the signal regions, \texttt{SR-lowMass} and \texttt{SR-highMass}, require exactly two taus with opposite-sign electric charge, and reject events with an additional third $\tau$ or light lepton. So only process in Eq.~(\ref{process4}) could contribute to the signal regions. The other searches corresponding to an integrated luminosity of 139 fb$^{-1}$ face the similar issues, such as requiring multiple jets~\cite{CMS:2019see,CMS:2019nvx,ATLAS:2019jvl}, hard jet~\cite{ATLAS:2019lov, ATLAS:2019nnv}, b-tagged jet~\cite{ATLAS:2019efx}, or exactly two light flavor leptons~\cite{ATLAS:2019lov}.

Given the fact that hundred of integrated luminosity 13 TeV events have been recorded at LHC, we further estimate the exclusion power of LHC with higher luminosity by normalizing signal and background event yields in the signal regions \texttt{SR-A44} and \texttt{SR-C18} of \cite{Sirunyan:2017lae}. We compute the significance as $\mathbf{S}=\sqrt{2(n_s+n_b)ln(1+n_s/n_b)-2n_s}$, where $n_s$ and $n_b$ are the normalized signal and background event yields, respectively. We show the result in the right panel of Fig.~\ref{lhc-R}. The samples with $m_H<$ 645 (700) GeV will be excluded at $2\sigma$ confidence level with 139 (300) fb$^{-1}$ integrated luminosity data. If the signal regions of \cite{Sirunyan:2017lae} are optimized for the production and decay modes in Eq. (\ref{process1}-\ref{decay-taumu}), the detection ability of LHC for this model could be further enhanced.

\section{Conclusion}
In this paper, we discuss a 2HDM in which one Higgs doublet has the same interactions with fermions as the SM, 
and another Higgs doublet only 
has the $\mu$-$\tau$ LFV interactions. Assuming the Yukawa matrices to be real and symmetrical, we 
considered various relevant theoretical and experimental 
constraints, and found that the excesses of muon $g-2$ and 
LUF in the $\tau$ decays can be simultaneously explained in many parameter spaces with 
300 GeV $<m_H<$ 800 GeV, $\Delta m<$ 50 GeV, and 0.36 $<\rho<$ 1.
The parameter spaces are sizable reduced by the direct search limits from the LHC,
and $m_H$ is required to be larger than 560 GeV.

{\em Note added:}
When this manuscript is being prepared, a similar paper appeared in the arXiv \cite{1907.09845}. 
Here we discussed different scenario, and obtain different conclusions.

\section*{Acknowledgment}
This work was supported by the Natural Science Foundation of
Shandong province (ZR2017JL002 and ZR2017MA004), by the National Natural Science Foundation
of China under grant 11575152, 11975013, and by the ARC Centre of Excellence for Particle Physics at the Tera-scale under the grant CE110001004.

\section*{Appendix}\label{app}
In this appendix, we explain in detail why the constraints from the ATLAS searches for two leptons and $E_T^{\rm miss}$~\cite{ATLAS:2019cfv} on Higgs masses in our model are weaker than that on slepton masses in the simplified model where only mass-degenerate $\tilde{e}$ and $\tilde{\mu}$ are considered. In Table II, we compare the cross sections, cut flows and event yield in each signal regions for a benchmark point in the simplified model, $(m_{\tilde{\ell}},m_{\tilde{\chi}_1^0})=(700,0)$ GeV, and a point in our model, $m_{H}=m_{A}=m_{H^{\pm}}=700$ GeV. 

We can see that the dominated process for our model are the Eq.~(\ref{process1}) and Eq.~(\ref{process3}), i.e. $pp\to H^{\pm}A,~H^{\pm}H$, and the cross section is larger than that of slepton pair production when $m_{H/A/H^{\pm}}=m_{\tilde{\ell}}$. However, for slepton pair production, the cut efficiency $\epsilon$ of each step is quit large, which means only small part of events are discarded. Furthermore, the surviving events are all assigned into same-flavour (SF) signal regions. While, for the Higgs production, requiring 2 opposite-sign (OS) leptons and number of jets $n_{\rm jets}$ less than 2 rejects most of the events, and the rest of events are separated into both SF and different-flavour (DF) signal regions. It is because that the final states of $H^{\pm}H(A)$ are either $1\mu+2\tau$ or $2\mu+1\tau$. For $1\mu+2\tau$ final state, when both of the $\tau$s decay hadronicly or leptonicly, the events will be rejected. The event with $2\mu+1\tau$ final state where $\tau$ decaying leptonicly also can not pass the 2 lepton requirement. Meanwhile, there is a possibility that the remaining two lepton are same-sign. Furthermore, the hadronic $\tau$ leads to an additional jet, which will decrease the possibility that the event passes the requirement of $n_{\rm jets}<2$ because there may be initial state radiation jet. The lepton from leptonic decay of $\tau$ can be $e$ or $\mu$, which means that the event will be allocated into SF or DF signal regions. As a result, the normalised events number $N$ in the most sensitive signal region, \texttt{SR-SF-1J} ($m_{T2}>160$~GeV), of slepton pair production is much larger than that of Higgs production in our model.

\begin{table}[th]\label{tab:compare}
\caption{Cross sections, cut flows and event yields in each signal region for a benchmark point of $(m_{\tilde{\ell}},m_{\tilde{\chi}_1^0})=(700,0)$ GeV in the simplified slepton model  and a point in our model with $m_{H}=m_{A}=m_{H^{\pm}}=700$ GeV. $\tilde{\ell}$ stands the first two generations of mass-degenerate sleptons. ``Combine" includes all non-SM Higgs productions in our model. $\epsilon$ indicates the cut efficiency and the normalised events number $N$ is the product of cross section,  integrated luminosity and $\epsilon$. The signal region name is abbreviated, such as ``DF-0J-100" standing \texttt{SR-DF-0J} ($m_{T2}>100$~GeV).}
\scriptsize
\begin{tabular*}{\hsize}{@{}@{\extracolsep{\fill}}|l||r|r||r|r||r|r|r|r|r|r|@{}}
\hline
Process                            & \multicolumn{2}{c||}{$\tilde{\ell}^{\pm}\tilde{\ell}^{\mp}$} & \multicolumn{2}{c||}{Combine}                               & \multicolumn{2}{c|}{$H^{\pm}A,~H^{\pm}H$}                          & \multicolumn{2}{c|}{$H^{\pm}H^{\mp}$}                      & \multicolumn{2}{c|}{$HA$}                                  \\ \hline
Cross section (fb)                 & \multicolumn{2}{c||}{0.178}                                  & \multicolumn{2}{c||}{0.336}                                 & \multicolumn{2}{c|}{0.218}                                 & \multicolumn{2}{c|}{0.064}                                 & \multicolumn{2}{c|}{0.054}                                 \\ \hline
\multicolumn{11}{|c|}{Cut Flow}                                                                                                                                                                                                                                                                                                                      \\ \hline
\multicolumn{1}{|c||}{}             & \multicolumn{1}{c|}{$\epsilon$}  & \multicolumn{1}{c||}{$N$} & \multicolumn{1}{c|}{$\epsilon$} & \multicolumn{1}{c||}{$N$} & \multicolumn{1}{c|}{$\epsilon$} & \multicolumn{1}{c|}{$N$} & \multicolumn{1}{c|}{$\epsilon$} & \multicolumn{1}{c|}{$N$} & \multicolumn{1}{c|}{$\epsilon$} & \multicolumn{1}{c|}{$N$} \\ \hline
No cut                             & 100\%                            & 24.8                     & 100\%                           & 46.7                     & 100\%                           & 30.3                     & 100\%                           & 8.9                      & 100\%                           & 7.5                      \\ \hline
2 OS leptons                       & 90.77\%                          & 22.5                     & 31.67\%                         & 14.8                     & 31.66\%                         & 9.6                      & 38.00\%                         & 3.4                      & 25.86\%                         & 1.9                      \\ \hline
$m_{\ell_1\ell_2}>100$ GeV         & 89.54\%                          & 22.2                     & 31.13\%                         & 14.5                     & 31.21\%                         & 9.5                      & 36.84\%                         & 3.3                      & 25.48\%                         & 1.9                      \\ \hline
No b-tagged jet                    & 83.75\%                          & 20.7                     & 28.34\%                         & 13.2                     & 28.37\%                         & 8.6                      & 34.54\%                         & 3.1                      & 22.54\%                         & 1.7                      \\ \hline
$E_T^{\rm miss}>110$ GeV           & 79.14\%                          & 19.6                     & 26.06\%                         & 12.2                     & 26.77\%                         & 8.1                      & 32.34\%                         & 2.9                      & 17.47\%                         & 1.3                      \\ \hline
$E_T^{\rm miss}$ significance$>10$ & 79.14\%                          & 19.6                     & 26.06\%                         & 12.2                     & 26.77\%                         & 8.1                      & 32.34\%                         & 2.9                      & 17.47\%                         & 1.3                      \\ \hline
$n_{\rm jets}<2$                   & 63.46\%                          & 15.7                     & 14.47\%                         & 6.8                      & 14.10\%                         & 4.3                      & 26.12\%                         & 2.3                      & 3.22\%                          & 0.2                      \\ \hline
$m_{\ell_1\ell_2}>121.2$ GeV\footnote{This cut is only applied to the events with two same-flavour leptons.}       & 62.96\%                          & 15.6                     & 14.38\%                         & 6.7                      & 14.04\%                         & 4.3                      & 25.87\%                         & 2.3                      & 3.21\%                          & 0.2                      \\ \hline
\multicolumn{11}{|c|}{Signal Regions}                                                                                                                                                                                                                                                                                                                \\ \hline
DF-0J-100                          & 0\%                              & 0                        & 0.76\%                          & 0.4                      & 0.49\%                          & 0.1                      & 2.27\%                          & 0.2                      & 0.05\%                          & 0.0                      \\ \hline
DF-0J-100-120                      & 0\%                              & 0                        & 0.04\%                          & 0.0                      & 0.03\%                          & 0.0                      & 0.20\%                          & 0.0                      & 0.01\%                          & 0.0                      \\ \hline
DF-0J-120-160                      & 0\%                              & 0                        & 0.13\%                          & 0.1                      & 0.08\%                          & 0.0                      & 0.40\%                          & 0.0                      & 0.01\%                          & 0.0                      \\ \hline
DF-0J-160                          & 0\%                              & 0                        & 0.58\%                          & 0.3                      & 0.38\%                          & 0.1                      & 1.67\%                          & 0.1                      & 0.03\%                          & 0.0                      \\ \hline
DF-1J-100                          & 0\%                              & 0                        & 1.67\%                          & 0.8                      & 2.08\%                          & 0.6                      & 1.52\%                          & 0.1                      & 0.31\%                          & 0.0                      \\ \hline
DF-1J-100-120                      & 0\%                              & 0                        & 0.14\%                          & 0.1                      & 0.16\%                          & 0.0                      & 0.13\%                          & 0.0                      & 0.02\%                          & 0.0                      \\ \hline
DF-1J-120-160                      & 0\%                              & 0                        & 0.26\%                          & 0.1                      & 0.32\%                          & 0.1                      & 0.29\%                          & 0.0                      & 0.06\%                          & 0.0                      \\ \hline
DF-1J-160                          & 0\%                              & 0                        & 1.27\%                          & 0.6                      & 1.60\%                          & 0.5                      & 1.10\%                          & 0.1                      & 0.23\%                          & 0.0                      \\ \hline
SF-0J-100                          & 32.89\%                          & 8.1                      & 2.68\%                          & 1.3                      & 1.24\%                          & 0.4                      & 10.43\%                         & 0.9                      & 0.17\%                          & 0.0                      \\ \hline
SF-0J-100-120                      & 1.07\%                           & 0.3                      & 0.13\%                          & 0.1                      & 0.06\%                          & 0.0                      & 0.45\%                          & 0.0                      & 0.03\%                          & 0.0                      \\ \hline
SF-0J-120-160                      & 2.39\%                           & 0.6                      & 0.26\%                          & 0.1                      & 0.15\%                          & 0.0                      & 0.93\%                          & 0.1                      & 0.03\%                          & 0.0                      \\ \hline
SF-0J-160                          & 29.43\%                          & 7.3                      & 2.29\%                          & 1.1                      & 1.04\%                          & 0.3                      & 9.05\%                          & 0.8                      & 0.12\%                          & 0.0                      \\ \hline
SF-1J-100                          & 23.48\%                          & 5.8                      & 5.95\%                          & 2.8                      & 6.60\%                          & 2.0                      & 7.05\%                          & 0.6                      & 1.72\%                          & 0.1                      \\ \hline
SF-1J-100-120                      & 0.72\%                           & 0.2                      & 0.24\%                          & 0.1                      & 0.34\%                          & 0.1                      & 0.29\%                          & 0.0                      & 0.11\%                          & 0.0                      \\ \hline
SF-1J-120-160                      & 1.64\%                           & 0.4                      & 0.57\%                          & 0.3                      & 0.69\%                          & 0.2                      & 0.71\%                          & 0.1                      & 0.25\%                          & 0.0                      \\ \hline
SF-1J-160                          & 21.13\%                          & 5.2                      & 5.13\%                          & 2.4                      & 5.58\%                          & 1.7                      & 6.06\%                          & 0.5                      & 1.36\%                          & 0.1                      \\ \hline
\end{tabular*}
\end{table}

\end{document}